\documentstyle[12pt,draft,aas_macros]{nature-ejb}
\font\large=cmbx10 scaled \magstep3

\newcommand{\be}{\begin{equation}}
\newcommand{\ee}{\end{equation}}

\newcommand{\ifm}[1]{\relax\ifmmode#1\else$\mathsurround=0pt #1$\fi}
\newcommand{\kms}{\ifmmode\,{\rm km}\,{\rm s}^{-1}\else km$\,$s$^{-1}$\fi}
\newcommand{\kpc}{\ifmmode\,{\rm kpc}\else kpc\fi}
\newcommand{\Mpc}{\ifmmode\,{\rm Mpc}\else kpc\fi}

\newcommand{\ltsima}{$\; \buildrel < \over \sim \;$}
\newcommand{\lsim}{\lower.5ex\hbox{\ltsima}}
\newcommand{\gtsima}{$\; \buildrel > \over \sim \;$}
\newcommand{\gsim}{\lower.5ex\hbox{\gtsima}}

\begin{document}

\title{\Large \bf Direct Formation of Supermassive Black Holes via Multi-Scale Gas
  Inflows in Galaxy Mergers}
\author{
  L. Mayer$^{1}$,
  S. Kazantzidis$^3$,
  A.Escala$^{4,5}$, 
  \& S. Callegari$^1$
\institute{$^1$Institute for Theoretical Physics, University of Z\"urich, Winterthurestrasse 190, 
8057 Z\"urich, Switzerland\\ 
$^3$Center for Cosmology and Astro-Particle Physics; 
        and Department of Physics; and Department of Astronomy, 
        The Ohio State University, USA\\
$^4$Kavli Institute for Particle Astrophysics and Cosmology (KIPAC), Stanford University, 2575 Sand Hill Road MS 29
Menlo Park, CA, USA\\
$^5$Departamento de Astronomia, Universidad de Chile, Casilla 36-D, 
Santiago, Chile.
}
}

\date{\today}{}
\headertitle{Direct supermassive black hole formation}
\mainauthor{Mayer et al.}

\summary{Observations of distant bright quasars suggest that billion solar mass supermassive
black holes (SMBHs) were already in place
less than a billion years after the Big Bang$^1$. Models in which light black hole seeds form 
by the collapse of primordial metal-free stars$^{2,3}$ cannot explain their rapid 
appearance due to inefficient gas accretion$^{4,5,6}$.  Alternatively, these
black holes may form by direct collapse of gas at the center of protogalaxies$^{7,8,9}$.
However, this requires metal-free gas that does not cool efficiently and thus is not turned into stars $^{8}$,
in contrast with the rapid metal enrichment of protogalaxies$^{10}$.
Here we use a numerical simulation to show that mergers between massive 
protogalaxies  naturally produce the required central gas accumulation with no need to
suppress star formation. 
Merger-driven gas inflows produce an unstable, massive
nuclear gas disk. Within the disk a second gas inflow accumulates more than
100 million solar masses of gas in a sub-parsec 
scale cloud in one hundred thousand years. The cloud 
undergoes gravitational collapse, which eventually leads to the formation
of a massive black hole. The black hole can grow to a billion solar 
masses in less than a billion years by accreting gas from the surrounding 
disk. 
}

\maketitle


The conventional scenario postulates that light black hole seeds form from the collapse of an early
generation of metal-free (Population III) stars$^{2,10,11}$.
Sustained accretion of such seed black holes at or above the Eddington rate is required to grow 
the billion solar masses SMBHs that are
thought to power bright quasars at $z \sim 6$$^3$. Yet numerical simulations show that the ionized gas
surrounding the seeds has very low densities and high pressures that would prevent the required high accretion rates$^4$ .
Radiative feedback from accretion and radiation pressure also  limit the accretion rate to values 
well below Eddington$^{5,6,12}$. 
Dynamical effects, such as the expulsion of a black hole from the host halo due to asymmetric emission
of gravitational waves in mergers (the "gravitational rocket")
can stifle further the growth of the seeds$^{3,13}$.

Alternatively, massive black hole seeds exceeding $10^4 M_{\odot}$ might form
directly from runaway gas collapse at the centers of protogalaxies$^{6,8,9,10,11,14}$.
The accumulation of large amounts of gas in galactic nuclei can occur via
efficient transport of angular momentum driven by gravitational torques in
galactic disks$^{15,16,17}$, via viscous diffusion due to gravito-driven
turbulence $^{18,19}$, or by means of the magnetorotational instability
$^{20,21}$. 
Recent models suggest that under normal
thermodynamical conditions of the interstellar medium
gas would be converted into stars faster than it is driven to the center$^{16,22}$. 
Instead, large 
inflow rates, $> 1 M_{\odot}$/yr, could accumulate a large central gas concentration
without fragmentation into stars, leading eventually to direct SMBH formation, if 
molecular cooling and metal  cooling are suppressed$^{8,16,22}$. 
However, even if molecular cooling can be suppressed at high redshift
as $H_2$, the most abundant molecule, is dissociated by the
cosmic ultraviolet background$^{8,16}$, as soon as some metals are produced after
the first generation of stars protogalaxies would still undergo rapid cooling and star formation, 
especially in the presence of dust$^{10}$.
Therefore, at present it is unclear if protogalaxies ever met the conditions required for direct
SMBH formation.

Yet, current direct formation models are simple one-dimensional 
semi-analytical calculations that start
from an initially stable, axisymmetric kiloparsec scale protogalactic gas disk, and perturb it under the
assumption that the inflow occurs in a steady-state fashion$^{8, 16}$.
Neither the initial stability
and regular structure, nor the steady state condition of the inflow, are representative
of galaxies at high redshift, which are subject to rapid gas accretion and repeated mergers. 
Moreover, due to their steady-state nature, these models cannot even demonstrate that
gas inflows do really lead to a central collapse.

The solution to  the rapid build-up of SMBHs may be offered by galaxy mergers.
In mergers tidal torques and shock dissipation are capable of driving most of the gas content of galaxies from
kiloparsec scales down to scales of several tens of parsecs at rates as high as
$10-100 M_{\odot}$/yr, forming nuclear gas
disks despite high star formation rates$^{23,24}$. If such gas inflows could continue all the
way down to the very center of the merger remnant at these high rates they would
provide an alternative route to direct massive black hole formation$^{10}$. 
Addressing this issue requires a three-dimensional simulation following gasdynamics 
across an unprecedented range of spatial scales.

\begin{figure}
\vspace{7cm} 
{\includegraphics{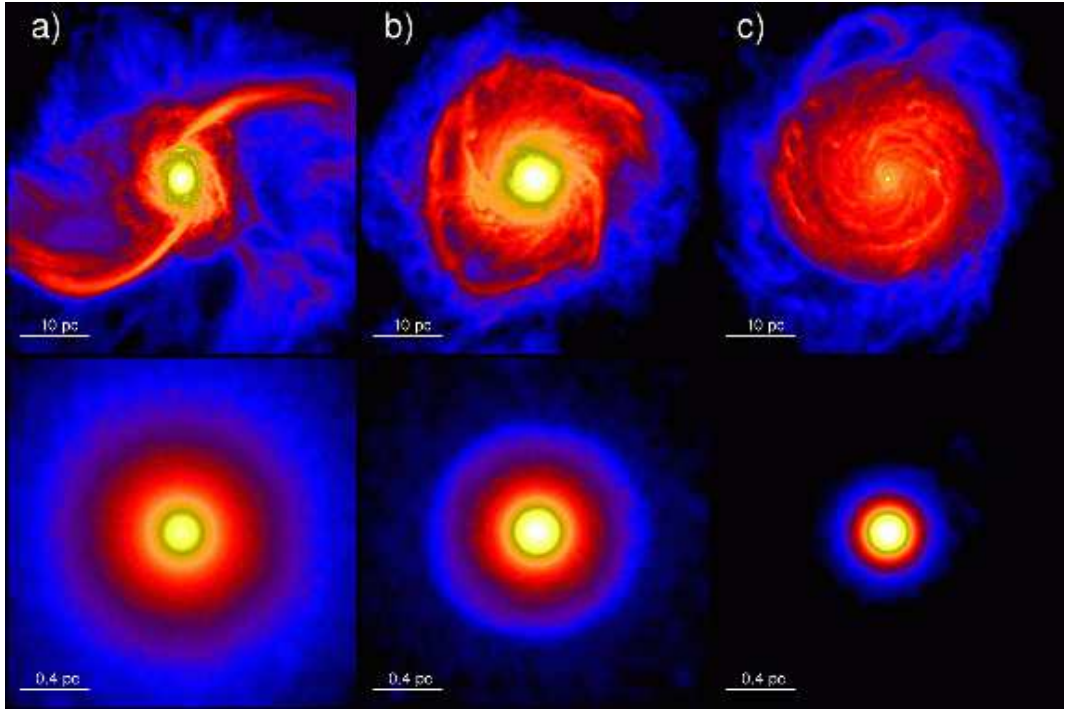}}
\caption[]{\small {\bf Time evolution of the simulated nuclear disk}. 
The surface density maps of the nuclear disk are shown at both large
(upper panels) and small scales (bottom panels). Particles are 
color-coded on a logarithmic scale with brighter colors in regions 
of higher density. The density ranges from $2 \times 10^4$ to 
$1 \times 10^8 M_{\odot}/pc^2$ and from $2 \times 10^6$ to 
$2 \times 10^{10} M_{\odot}/pc^2$ in the upper and lower panels,
respectively. The time increases from left to right, corresponding to 
$9.1 \times 10^3$ yr, $7.49 \times 10^4$ yr and $1.036 \times 10^5$ yr after the merger. 
The time of the merger is defined as the time at which the two density peaks associated
with the merging galactic cores are no longer distinguishable.
For reference, the disk orbital time at $\sim 20$ pc is $5 \times 10^4$ yr,
while at $1$ pc it is $\sim 4 \times 10^3$ yr. 
Global spiral modes, in particular the two-armed spiral initially triggered
by the final collision between the two cores, are evident at scales of tens of
parsecs (upper panels) and cause the mass increase in the central parsec
region (bottom panels (a)-(b)) that allows the collapse into a massive central 
cloud (from bottom panel (b) to bottom panel (c). 
}
\end{figure}

We begin by performing a merger simulation between two identical disk galaxies with moderate amounts of 
gas$^{23}$, several kiloparsecs in size and embedded in a $10^{12} M_{\odot}$ halo
(see Supplementary Information). In the concordance WMAP5 cosmology such objects 
would correspond to fairly high density peaks collapsing at $z \sim 7-8$$^5$, ($\sim
4-5\sigma$, where $\sigma$ is the rms variation of the cosmological density field).
Recent cosmological simulations show that massive, kiloparsec-scale 
rotating disks of stars and gas are already present in halos with masses 
$> 10^{11} M_{\odot}$ at $z > 4$$^{21}$. The mass of the dark halo of
each galaxy is consistent with that
inferred for the hosts of high-redshift quasars based on their number
densities$^{5,25}$. The two galaxies are placed on a parabolic orbit whose
parameters are consistent with those found in cosmological simulations$^{23,24}$.
We employ the technique of smoothed particle
hydrodynamics (SPH) to model the galaxy collision, including
radiative cooling, star formation and a temperature floor at $2
\times 10^4$ K to mimic non-thermal pressure due to turbulence in the
interstellar medium (ISM) (see Supplementary Information). 

The two merging galaxies undergo two close encounters as dynamical friction
against their extended halos erodes their orbital energy. At the time when the 
two baryonic cores are separated by 6 kpc, when they begin their last orbit
before the final collision, we perform particle splitting in the gas component within a volume 
30 kpc in size (See Supplementary Information). As a result, we increase our gas mass resolution
eight-fold and, simultaneously, we decrease the gravitational
softening achieving a spatial resolution of $0.1$ pc (see Supplementary Information). 
In this refined calculation we adopt an effective equation of state (EOS) that
accounts for the net balance between cooling and heating$^{26}$. A lower
resolution un-refined simulation shows that during the final collision
the star formation peaks at $\sim 30 M_{\odot}$/yr over about $10^8$yr$^{23}$.
Motivated by this, we chose an EOS with an effective $\gamma$ in the 
range $1.1-1.4$ which is appropriate for interstellar gas heated 
by a major starburst$^{27}$ and assuming solar metallicity as suggested by the
abundance analysis of the high-redshift QSOs$^{22}$ (see Supplementary Information).

The final collision of the two galactic cores produces a massive turbulent
rotating nuclear disk with a mass of $\sim 2 \times 10^9 M_{\odot}$ and a
radius of about 80 pc$^{24}$. With the increased resolution we can follow the
internal evolution of the nuclear disk.
The disk is born in an unstable configuration, with a prominent two-armed spiral pattern imprinted
by the collision and sustained by its
own strong self-gravity (Figure 1). The gas has a high turbulent velocity
dispersion ($\sigma \sim 100$ km/s) maintained by gravitational instability$^{18}$ and rotates at a speed of several hundred km/s within 50 pc. It constitutes most of the mass in the nuclear
region, the rest being in dark matter and stars. The spiral arms 
swiftly transport mass inward and angular momentum outward $^{28}$.
About only $10^4$ yr after the merger is completed, more than $20\%$ of the disk 
mass ($> \sim 5 \times 10^8 M_{\odot}$) resides within the central few parsecs
(Figure 2) where the inflow rate peaks at $\dot{M} > 10^4 M_{\odot}$/yr (see 
Figure 1 of Supplementary Information) This corresponds
to an inflow timescale $t_{\rm inf} = M/\dot{M} < T_{\rm orb}$, where $T_{\rm 
  orb}$ is the local orbital timescale. The inflow rate is orders of
magnitude higher than that expected in isolated, thin locally gravitationally 
unstable disks considered in analytic models of direct massive black hole 
formation$^{16}$, but it is consistent with results of three-dimensional 
simulations of disks subject to global instabilities$^{18,28}$ 
(see Supplementary Information). The self-gravitating nuclear disk does not fragment because 
of its high effective thermal pressure and even higher turbulent pressure, which maintain
a Toomre Q parameter above the threshold for stability (see Supplementary Information).

The gas funnelled to the central 2-3 pc region of the nuclear disk settles into 
a rotating, pressure supported cloud.
The density of the cloud continuously increases as it gains mass from the
inflow until it becomes Jeans unstable and collapses to sub-parsec scales on
the local dynamical time, $t_{\rm dyn} \sim 10^3$ yr (Figure 1). The
supermassive cloud contains $\sim 13 \%$ of the disk mass ($\sim 2.6 \times
10^8 M_{\odot}$) (Figure 2). The simulation is stopped once the central cloud
has contracted to a size comparable to the spatial resolution limit. 
At this point the cloud is still Jeans unstable.
With greater resolution its collapse should
continue since the equation of state would become essentially isothermal
at even higher densities$^{27}$. With its steep density profile ($r \sim \rho^{-\gamma}$, with
$\gamma > 2$) the cloud would be very massive and dense even at much smaller radii. 
At a radius $\sim 10^{-3}$ pc it would match the conditions of 
a quasi-star that can then collapse directly into a massive black hole$^{9,10}$.
Alternatively, it could collapse on a free-fall timescale into a massive black hole
without prior formation of a quasi-star.
Only $\sim 10^5$ yr have elapsed since the completion of
the merger, a timescale much shorter than the $10^8$ years needed to convert
most of the nuclear gas into 
stars during the starburst (see Supplementary Information). Therefore, in  our
merger simulation the gas flows inward and collapses into a compact object much 
faster than it can form stars,
overcoming the major difficulty of previous direct formation models appealing
to metal-free conditions in protogalaxies to suppress star formation$^{10,16}$. 

The very high temperature of the cloud, $T > 10^7$ K, makes star formation in
its interior highly unlikely. Yet, even if less 
than $1\%$ of the cloud mass collapsed into a black hole
it would still produce an object of mass $> 10^5 M_{\odot}$ which can
subsequently grow by accreting the surrounding nuclear gas disk (see Supplementary
Information). Assuming that
the disk forms stars with a $\sim 30\%$ efficiency,
as deduced from observation and models of star forming molecular clouds$^{27}$, a
gas mass in excess of $10^9 M_{\odot}$ would still be available to feed the
black hole over a timescale longer than $10^8$ yr, the duration of the starburst. 
Considering Eddington-limited accretion$^{29}$ the black hole can 
grow to $M_{BH} \sim 10^9 M_{\odot}$  in as little as $3.6 \times 10^8$ yr (see
Supplementary Information).
Therefore the massive seed black hole
can grow fast enough for bright quasars to arise within the first billion year from the Big Bang, namely by $z \sim 6$. This provided that 
the galaxy merger also occurs on a timescale shorter than a billion year, a natural condition in CDM models at high redshift due
to the high densities and short orbital timescales of collapsed objects$^{30}$ (see Supplementary Information).

\begin{figure}
\vskip 6cm
{\includegraphics{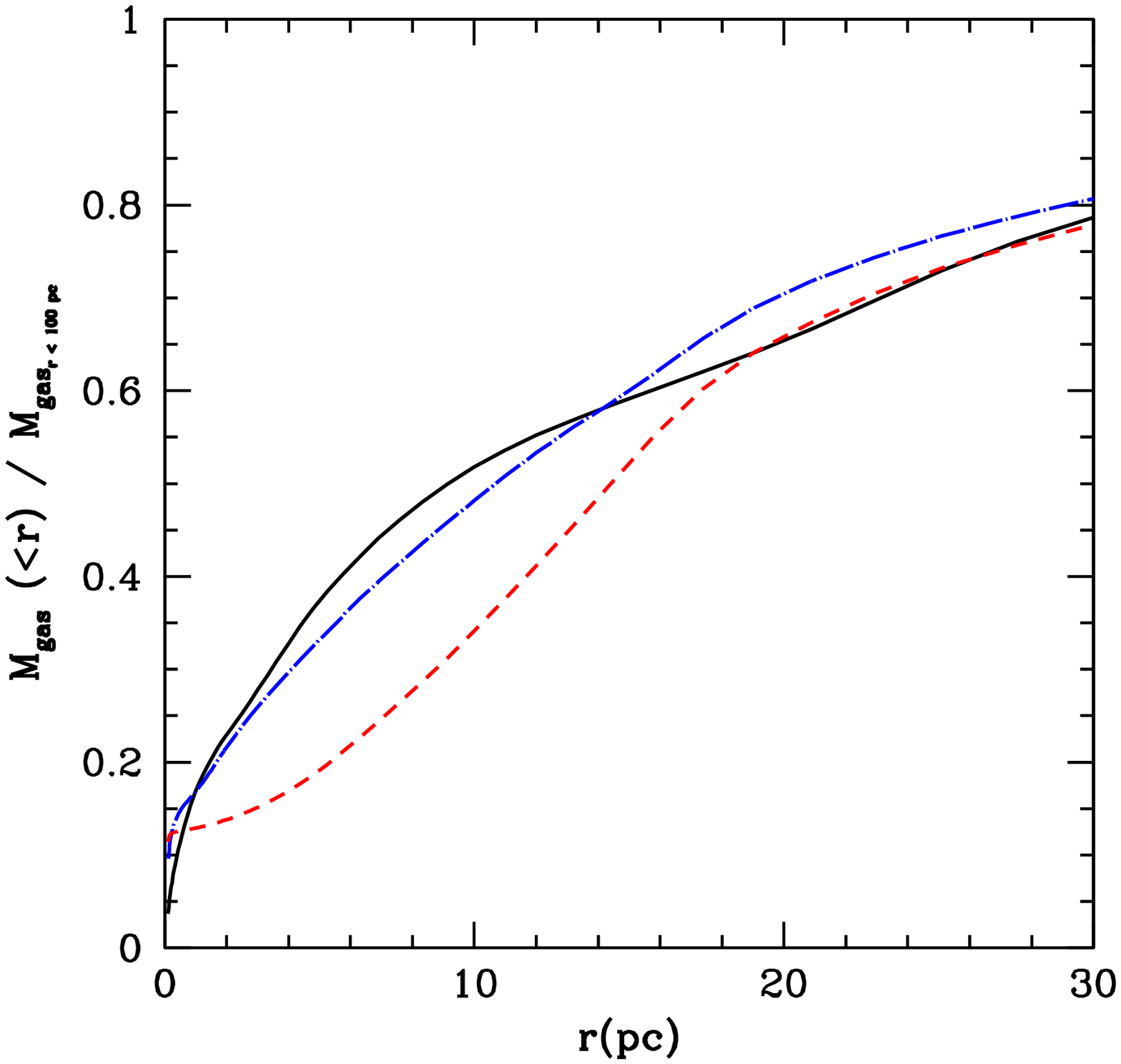}} 
{\includegraphics{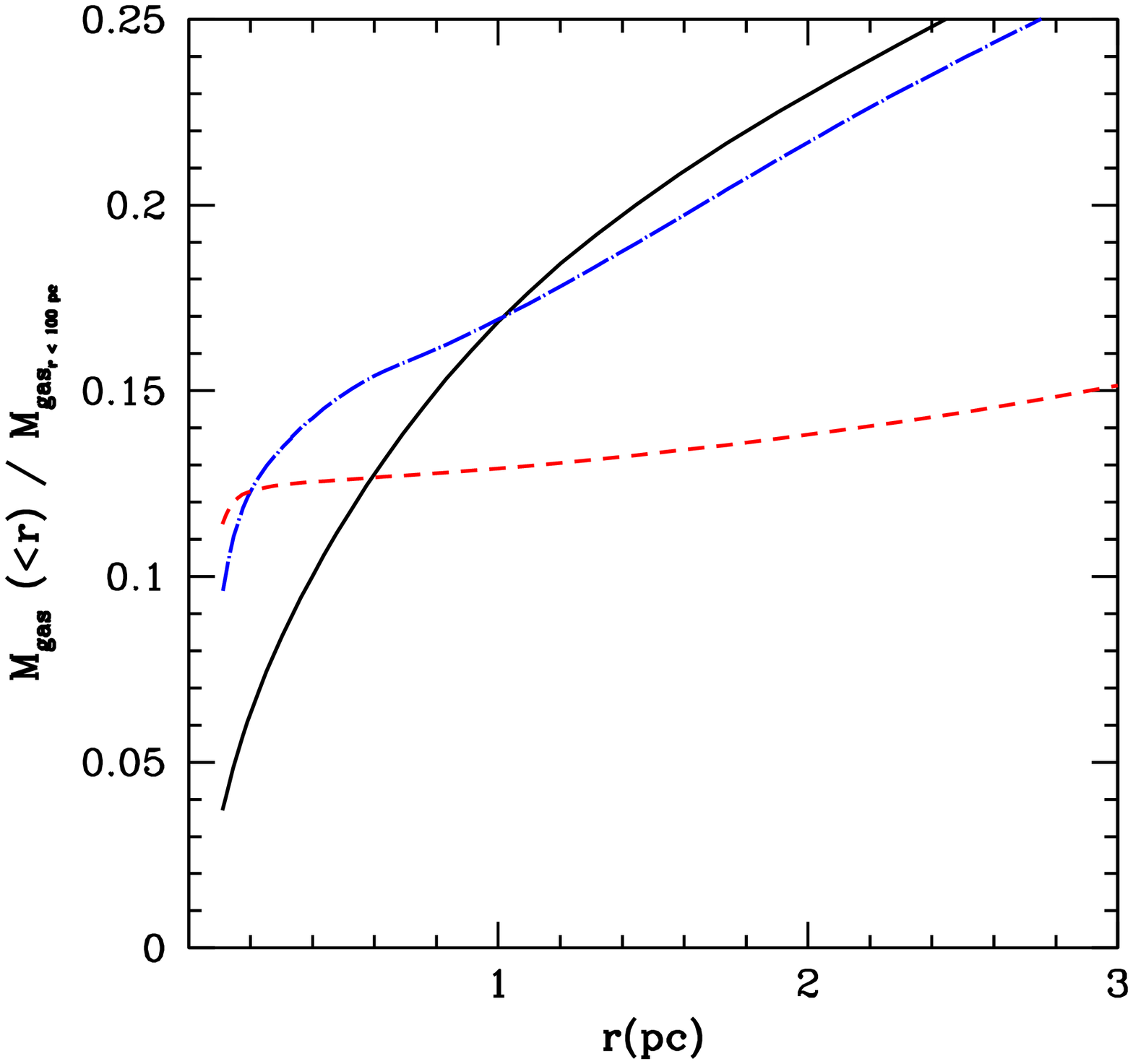}}
\caption[]{\small {\bf Evolution of the cumulative gas mass profile of the nuclear disk}. Profiles are 
  normalized to the total gas mass within the radius the nuclear disk (100 pc). The
disk radius is determined from the sharp drop in the nuclear gas density distribution.
The left panel shows the profile at scales of tens of parsecs, while the right
panel displays the profile within the inner few parsecs. In both panels
different curves show the mass profile at $9.1 
\times 10^3$ yr (solid line), $7.49 \times 10^4$ yr 
(dot-dashed line) and  $1.036 \times 10^5$ yr (dashed line)
after the merger, these being the same snapshots used in Figure 1.
The time of the merger is defined as the time at which the two density peaks associated
with the merging cores of the galaxies are no longer distinguishable. Mass redistribution is evident as spiral arms 
at tens of parsecs
scales push mass inward and shed angular momentum outwards (top panel), gradually leading to an increasing mass 
concentration
in the central region (bottom panel). This triggers the 
Jeans collapse of the central inner few parsecs into a supercloud containing $ \sim 13\%$ of the total disk mass, 
which manifests
as a strong flattening of the profile at parsec scales as the cloud absorbs a large fraction of the mass in that region}
\end{figure}



Large deviations from the local 
$M_{\rm BH}- \sigma$ relation (where $\sigma$ is the velocity dispersion 
of the central stellar spheroid)$^{22}$ should be expected at high redshift as the
black hole mass grows faster than that of its host galaxy.
A black hole mass much larger than that inferred from the local $M_{\rm BH} - \sigma$ 
relation has actually been suggested for the only high-redshift quasar for
which gas kinematics has been measured$^{25}$. Substantial evolution in the
high-redshift scaling relations between SMBHs and their host 
galaxies will be testable with future observations of bright quasar hosts 
with the James Webb Space Telescope (JWST) and the Atacama Large Millimeter
Array (ALMA).  At low redshift the formation mechanism that we propose should
be suppressed as pre-existing SMBHs in the two merging galaxies heat the surrounding
medium while they accrete gas, preventing instabilities and inflows in the nuclear
disk. Finally, future gravitational wave experiments such as 
the Laser Interferometer Space Antenna (LISA) 
will be able to test the existence of a population of black holes that
had a jump start by probing the mass distribution of merging black holes
as a function of redshift .


\begin{acknowledge}
Stimulating discussions with Monica Colpi, Richard Durisen, Fabio Governato, Piero Madau,
Thomas Quinn, David Weinberg, Romain Teyssier and Simon White are
greatly acknowledged. This research has
been supported by the Swiss National Science Foundation (SNF), 
by the Center for Cosmology and  Astro Particle Physics (CCAPP) at Ohio State University, and by the
Kavli Institute for Particle Astrophysics (KIPAC) 
at Stanford University. LM, SK and SC aknowledge the Kavli Institute for Theoretical Physics at the University
of California in Santa Barbara (KITP)
for hospitality during the initial stages of this work during the program ``Building the Milky Way''.
All computations were performed on the Zbox3 supercomputer at the University
of Z\"urich and on the Brutus cluster at ETH Z\"urich.
\end{acknowledge}

\vskip 0.5cm
{\noindent {\bf Author Information} The authors declare that they have no competing financial
interests. Correspondence and requests for materials should be addressed to L.M. 
(lucio@phys.ethz.ch) or S.K. (stelios@mps.ohio-state.edu).

\newpage


\begin{center}
\vspace*{-10.00pt}
{\Large \bf SUPPLEMENTARY INFORMATION}
\end{center}

\noindent Here we briefly describe the setup of the initial conditions and the
numerical methods used to perform the simulations presented
in the Letter.
This is followed by a critical discussion of the assumptions behind
the modeling of thermodynamics in the simulations and by a quantitative
discussion of the instability and dynamics of the nuclear disk to
support the results described in this Letter. We conclude with a discussion
of the growth of the massive black hole seed using analytical estimates.

\section{Numerical Methods}

\subsection{The N-Body+SPH code:GASOLINE}

We have used the fully parallel, N-Body+smoothed particle hydrodynamics (SPH)
code GASOLINE to compute the evolution of both the collisionless and
dissipative component in the simulations. A detailed description of the code
is available in the literature$^1$. Here we recall its essential features.
GASOLINE computes gravitational forces
using a tree--code$^2$ that employs multipole expansions to approximate the gravitational
acceleration on each particle. A tree is built with each node storing
its multipole moments.  Each node is recursively divided into smaller
subvolumes until the final leaf nodes are reached.  Starting from the
root node and moving level by level toward the leaves of the tree, we
obtain a progressively more detailed representation of the underlying
mass distribution. In calculating the force on a particle, we can
tolerate a cruder representation of the more distant particles leading
to an $O(N \log{N})$ method. Since we only need a crude representation
for distant mass, the concept of ``computational locality'' translates
directly to spatial locality and leads to a natural domain decomposition.
Time integration is carried out using the leapfrog method, which is a
second-order symplectic integrator requiring only one costly force
evaluation per timestep and only one copy of the physical state of the system.

SPH is a technique of using particles to integrate
fluid elements representing gas$^{3.4}$.
GASOLINE is fully Lagrangian, spatially and temporally adaptive and efficient
for large $N$. It employs
radiative cooling in the galaxy merger simulation used as a starting point for the refined
simulations presented in this Letter. We use a standard cooling
function for a primordial mixture of atomic hydrogen and helium. We shut off radiative
cooling at temperatures below $2 \times 10^{4}$ K that is
about a factor of $2$ higher than the temperature at which atomic radiative
cooling would drop sharply due to the adopted cooling function.
With this choice we take into account non-thermal, turbulent pressure to model
the warm ISM of a real galaxy$^{5}$. Unless strong shocks occur (this will be
the case during the final stage of the merger) the gaseous disk evolves nearly isothermally
since radiative cooling is very efficient at these densities ($< 100$ atoms/cm$^3$) and
temperatures ($10^4$ K), and thus dissipates rapidly the compressional heating resulting
from the non-axisymmetric structures (spiral arms, bars) that soon develop in each
galaxy as a result of self-gravity and the tidal disturbance of the companion.
The cooling rate would increase with the inclusion of metal lines, but it
has been shown (see section 1.4) that the equation of state of gas at these densities is still nearly isothermal
($\gamma \sim 0.9-1.1$) for a range of metallicities (with $\gamma$ being lower for higher
metallicity), supporting the validity of our simple choice for the cooling function.
Cooling by metals will surely be important below $10^4$ K, but this would be irrelevant
in our scheme since we have imposed a temperature floor of $2 \times 10^4$ K to account
for non-thermal pressure (see above). The specific internal
energy of the gas is integrated using the asymmetric formulation. With this formulation
the total energy is conserved exactly (unless physical dissipation due to cooling
processes is included) and entropy is closely conserved away from shocks,
which makes it similar to alternative entropy integration approaches$^{6}$.
Dissipation in shocks is modeled using the
quadratic term of the standard Monaghan artificial viscosity$^{4}$.
The Balsara correction term is used to reduce unwanted shear viscosity$^{7}$.
The galaxy merger simulation$^{8}$ includes star formation as well.
The star formation algorithm is such that gas particles in dense, cold Jeans unstable regions and in
convergent flows spawn star particles at a rate proportional to the local dynamical
time$^{9,10}$. The star formation efficiency was set to $0.1$,
which yields a star formation rate of $1-2 M_{\odot}$/yr for models in isolation that have a
disk gas mass and surface density comparable to those of the Milky Way.

\subsection{The simulations of galaxy mergers}

For the Letter we performed a refined calculation of a galaxy merger simulation
between two identical galaxies.
The initial conditions of this and other similar merger simulations are described in 
previous papers$^{8,9}$.
The models are based on our knowledge of present-day disk galaxies simply because
there is little information available from obervations of galaxy structure at
high redshift. Yet, this modeling strategy should be regarded as conservative for the purpose
of this paper since high redshift disks are denser, more gas-rich and more turbulent
than present-day galaxies, which should favour the formation of massive
nuclear disks and, subsequently, of large gas concentrations within them.
We employed a multicomponent galaxy model constructed using the technique originally developed
in $^{11,12}$, its structural parameters being consistent with the $\Lambda$CDM paradigm 
for structure formation$^{13}$.
The model comprises a spherical and isotropic Navarro-Frenk-and-White (NFW)
dark matter (DM) halo$^{14,15}$, an exponential disk, and
a spherical,  non-rotating bulge. 
We adopted parameters
from the Milky Way model A1 of $^{16}$. Specifically, the DM
halo has a virial mass of $M_{\rm vir}=10^{12} M_{\odot}$, a
concentration parameter of $c=12$, and a dimensionless spin parameter
of $\lambda=0.031$. The mass, thickness and resulting scale length of the disk
are $M_{\rm d}=0.04 M_{\rm vir}$, $z_{0}=0.1 R_{\rm d}$, and $R_{\rm d}=3.5$ kpc,
respectively. The bulge mass and scale radius are $M_{\rm b}=0.008 M_{\rm vir}$
and $a=0.2 R_{\rm d}$, respectively.  While a stellar bulge is always present in
the most massive disk galaxies out to $z = 1$$^{17}$ there is still insufficient
knowledge of galactic structure at higher redshift to confirm that this is the case
also at $z > 6$. However, the presence of the bulge has the effect of stabilizing the galaxy disks 
against external perturbations$^{18}$. This implies that, without the bulge,
the two disks would undergo even stronger non-axisymmetric distortions during the final stage
of the merger and initiate a gas inflow even more powerful that the one obtained in our
current calculation. Hence including the bulge should be regarded as a conservative
assumption for the purpose of this Letter.

The DM halo was adiabatically contracted
to respond to the growth of the disk and bulge$^{19}$ resulting
in a model with a central total density slope close to isothermal.
The galaxies are consistent with the stellar mass Tully-Fisher and size-mass relations.
The gas fraction, $f_{\rm g}$, is 10\% of the total disk mass. This is fairly typical
for Milky Way-sized galaxies at low redshift but it is a conservative assumption for 
galaxies at $z > 2$$^{20}$.
The rotation curve of the model is shown in Figure 1.
The simulation
presented in this Letter is the refined version of a coplanar prograde encounter. This
particular choice is by no means special for our purpose, except that the galaxies
merge slightly faster than in the other cases, thus minimizing the computational
time invested in the expensive refined simulation. In particular, in previous work 
we have shown that the existence
of a coherent nuclear disk after the merger is a general result that does not depend
on the details of the initial orbital configuration, including the initial relative inclination
of the two galaxies$^{8,9}$. Similarly, gas masses and densities
in the nuclear region were found to differ by less than a factor of 2 for runs having the same initial
gas mass fraction in the galaxy disks but different initial orbits.


The galaxies approach each other on parabolic orbits with pericentric distances
that were $20\%$ of the galaxy's virial radius, typical of cosmological
mergers$^{21}$ . The initial separation of the halo centers was twice
their virial radii and their initial relative velocity was determined from the corresponding
Keplerian orbit of two point masses. Each galaxy consists of $10^5$
stellar disk particles, $10^5$ bulge particles, and $10^6$ DM particles.
The gas component was represented by $10^5$ particles.
We adopted a gravitational softening of $\epsilon = 0.1$ kpc
for both the DM and baryonic particles of the galaxy.

\subsection{The refined simulations of the nuclear region}

\subsubsection{Particle splitting}

In this Letter we use the same technique of static particle splitting that has been used before to study the dynamics of supermassive
black hole binaries evolving in circumnuclear gaseous disks$^{22}$ as well as
during galaxy mergers$^8$, and to study the assembly of galaxies from the cooling flow in a galaxy-sized halo$^{23}$. 
A similar technique has been used by others to 
study the dynamics of binary black holes in spherical gaseous backgrounds$^{24}$.
In dynamic splitting the mass resolution is increased during the simulation based on some criterion,
such as the local Jeans length of the system. This requires extreme care when calculating SPH density or pressure at the boundary 
between the fine grained 
and the coarse grained volumes$^{25}$.
In static splitting, the approach is much more  conservative and one simply selects a subvolume to refine.
The simulation is
then restarted with increased mass resolution just in the region of interest.
By selecting a large enough volume for the fine grained region one can
avoid dealing with spurious effects at the coarse/fine boundary.
We select the volume of the fine-grained region large enough to guarantee that the dynamical timescale
of the entire coarse-grained region is much longer than the dynamical timescale of the refined region.
In other words, we make sure that gas particles from the coarse region will reach the
fine region on a timescale longer than the actual time span probed in this work. This is important because
the more massive gas particles from the coarse region can exchange energy with the lower mass particles
of the refined region via two-body encounters, artificially affecting their dynamics and thermodynamics$^{26}$.
Hence our choice to split in a volume of 30 kpc in radius, while the two galaxy cores are separated
by only $6$ kpc.
The new particles are randomly distributed according to the SPH smoothing kernel within a volume  of size 
$\sim h_p^3$, where $h_p$ is the smoothing 
length of the parent particle. The velocities of the child particles are equal to those of their parent particle (ensuring 
momentum conservation)
and so is their temperature, while each child particle is assigned a mass equal to $1/N_{\rm split}$ the mass of the parent 
particle, where
$N_{\rm split}$ is the number of child particles per parent particle.
The mass resolution in the gas component was originally $2 \times 10^4 M_{\odot}$
and becomes $\sim 3000 M_{\odot}$ after splitting, for a total of 1.5 million SPH
particles.. The star and dark matter particles are not splitted
to limit the computational burden.
The softening of the gas particles is reduced to $0.1$ pc (it was $100$ pc in the low resolution simulations).
For the new mass resolution, the local Jeans length is always resolved by 10 or more SPH smoothing
kernels$^{27,28}$ in the highest density regions occurring in
the simulations. 

The softening of dark matter and star particles remains $100$ pc because these are not splitted. Therefore
in the refined simulations stars and dark matter particles essentially provide a smooth background
potential to avoid spurious two-body heating against the much lighter gas particles, while the computation focuses on the 
gas component which dominates by mass in the nuclear region.
By performing numerical tests we have verified that, owing to the fact that gas dominates the mass and dynamics of the
nuclear region, the large softening adopted for the dark matter particles does not affect
significantly the density profile of the inner dark halo that surrounds the nuclear disk.

\subsection{Thermodynamics of the nuclear region}

\paragraph{Model description}

In the refined simulations the gas is ideal and each gas particle obeys $P=(\gamma - 1) \rho u$.
The specific internal energy $u$ evolves with time as a result of $PdV$ work and shock heating modeled via
the standard Monaghan artificial viscosity term (no explicit radiative cooling term is included). 

The entropy of the system increases as a result of shocks. Including irreversible heating from shocks is
important in these simulations since the two galaxy cores undergo a violent collision.
Shocks are generated even later
as the nuclear, self-gravitating disk becomes non-axisymmetric, developing strong spiral arms.
Therefore the highly dynamical regime modeled here is much different from that considered by previous works
starting from an equilibrium disk model, which could be evolved using a polytropic equation of state and
neglecting shock heating$^{22,29}$.
Radiative cooling is not directly included in the refined simulations. Instead, the magnitude of the
adiabatic index, namely the ratio between specific heats, is changed in order to mimic different
degrees of dissipation in the gas component, thereby turning the equation of state of the
gas into an ``effective'' equation of state$^{30,31}$. We have shown elsewhere$^{9}$ that
the transition between the radiative cooling regime and the effective equation of state regime
does not introduce numerical artifacts in the simulation.

Previous works$^{31}$ have used a two-dimensional radiative transfer code to study the
effective equation of state of interstellar clouds exposed to the intense UV radiation field
expected in a starburst finding that the gas has an adiabatic index $\gamma$
in the range $1.1-1.4 (=7/5)$ for densities in the range $5 \times 10^3-  10^5$ atoms/cm$^3$,
comparable to the volume-weighted mean density in the simulated nuclear
disk. Such values of the adiabatic index are expected for quite a range of starburst intensities, from
$10$ $M_{\odot}$/yr to more than $100$ $M_{\odot}$/yr, hence encompassing the peak star formation 
rate of $\sim 30 M_{\odot}$/yr measured in the original low-res galaxy merger simulations$^{8}$.
Hence under these conditions the nuclear gas is not isothermal ($\gamma=1$),
which would correspond to radiative cooling being so efficient to balance heating coming from 
compression and/or radiative processes, as it happens in the first stage of the simulation.
Its inefficient cooling at densities of $10^4$ atoms/cm$^3$ is mostly due to a high optical depth which causes
trapping in H$_2$O and CO lines. In addition the warm dust heated by the starburst continuously heats the gas
via dust-gas collisions, and  the cosmic-rays do so as well$^{31}$.
We adopt $\gamma = 7/5$ until the first gas inflow is completed in the simulation (we have
also run a case for $\gamma=1.3$ and found that the structure of the nuclear disk is substantially
unchanged). 
That the mean properties (mass, density, pressure support contributed
by the thermal and turbulent components, rotational velocity)  of nuclear disks formed in 
galaxy mergers$^{32}$ and simulated with such an effective equation of state compare well with the 
corresponding properties of nuclear disks observed in detailed observations of merger remnants
has been already shown in previous work$^{9}$.
For densities above $10^5$ atoms/cm$^3$ cooling is more efficient and
$\gamma$ should drop to $\sim 1.1$ according to the adopted EOS model$^{31}$. This condition is verified in the central few parsecs
after the first inflow ($\sim 10^4$ yr after the merger), hence $\gamma = 1.1$ is adopted from this
time onward. By comparing with another simulation that was continued with $\gamma=1.4$ we verified that the 
dynamics of the disks at larger scales (tens of parsec scales) are not affected by varying $\gamma$.

With our EOS we model the nuclear gas as a one-phase medium.
In reality the nuclear disks will have a complex multi-phase structure with temperatures
and densities spanning orders of magnitude even in localized regions, as shown by detailed numerical 
calculations of nuclear disk models$^{33,34}$.
In particular, even the same radius the disk may have regions with different densities that may evolve
as if the equation of state was varying locally. Furthermore, based on the assumed EOS model$^{31}$ the lowest density ($ < 10^3$ atoms/cm$^3$) 
and highest density gas ($ > 10^6$ atoms/cm$^3$) is characterized 
by an adiabatic index $\gamma \leq 1$, hence by
a lower sound speed $v_s = \sqrt{\gamma k_B T / \mu}$.
A lower gas sound speed will make
the disk more gravitationally unstable, lowering the Toomre Q parameter, and thus more prone to fragmentation
and star formation$^{35}$.  However, as the Toomre parameter is lowered stronger asymmmetries
and spiral modes will also occur, increasing the gas inflow. The question is then whether the net amount of mass
transported to the center will also increase or will be reduced by the competing effect of star
formation. Recent numerical work$^{36}$ has shown that even with specific star formation rates 5 times higher than
expected based on the Kennicutt-Schmidt law gas inflows are still prominent and lead to a significant
deposition of mass to the center of an isolated rotating nuclear gas disk model, strongly suggesting
that the results presented in this Letter will remain valid even in a more realistic calculation
incorporating star formation and local variations of the effective equation of state. 

\begin{figure}
\vskip 11.6cm 
{\includegraphics{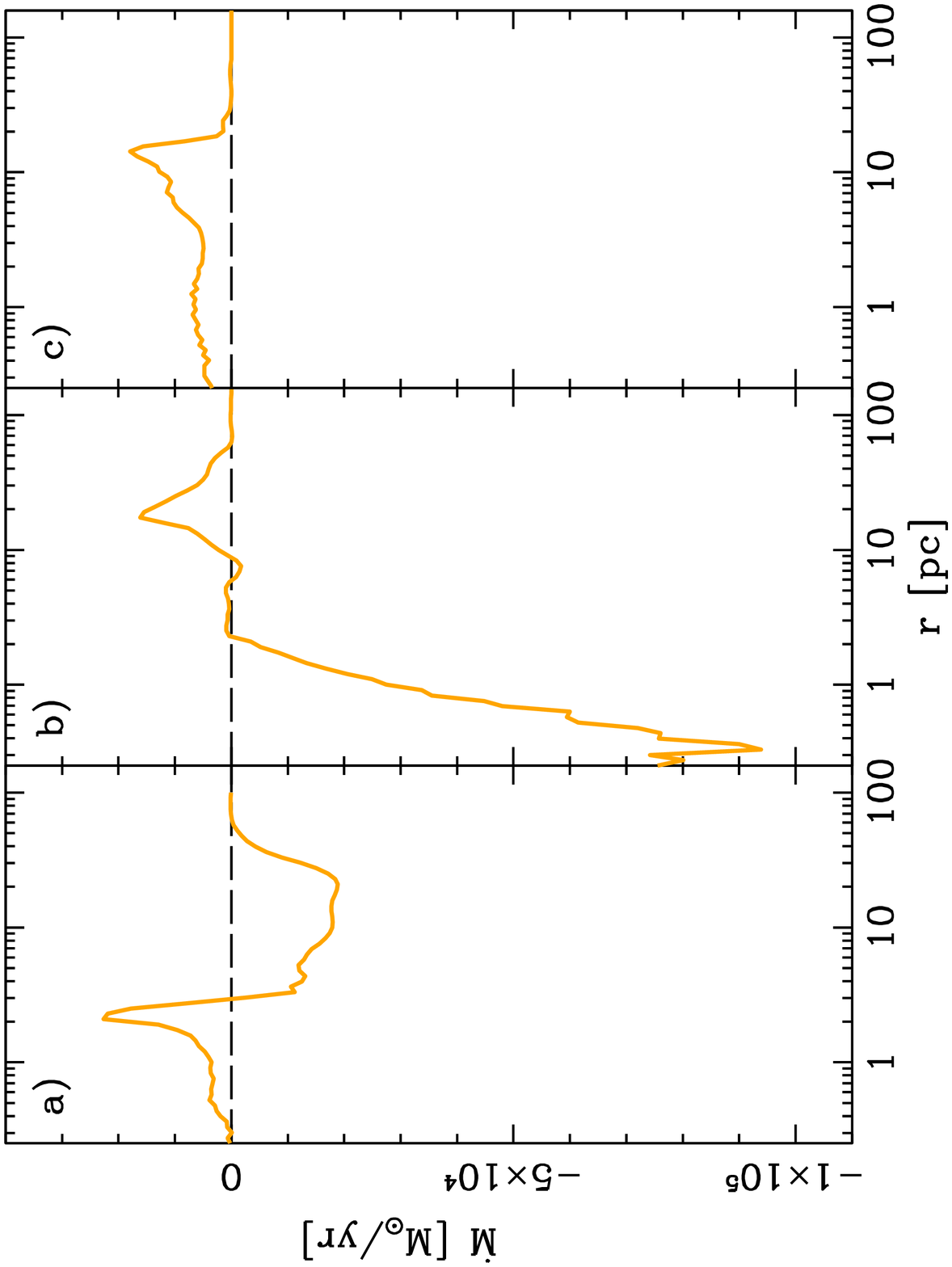}}
{\includegraphics{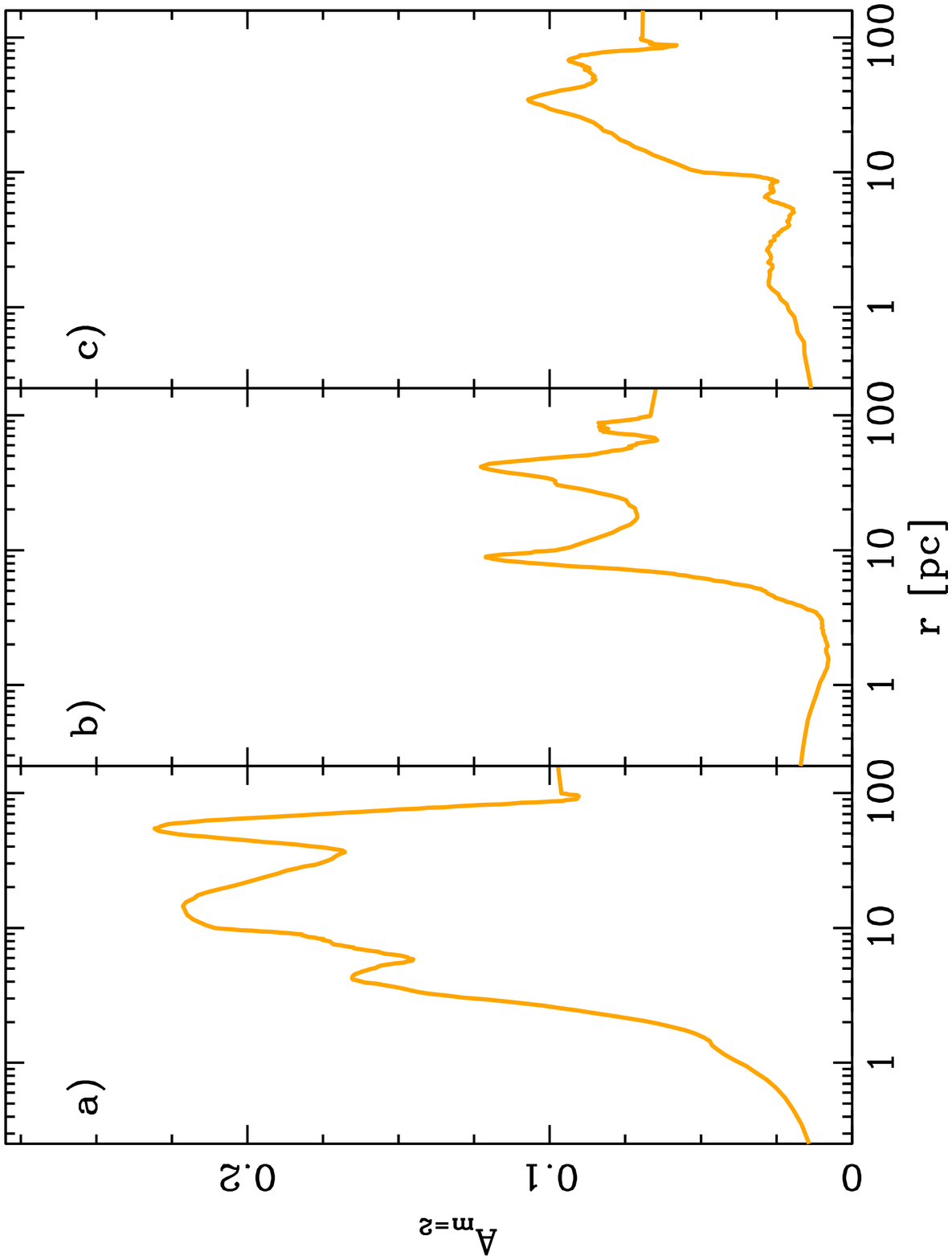}}
\caption[]{\small 
Time evolution of the gas inflow rate (top panel) and of the amplitude of the strongest non-axisymmetric mode in the disk (bottom panel, 
$m=2$ is the strongest mode at all times). Azimuthally averaged radial profiles are shown in both cases, time averaged over a few outputs
around a chosen time $t_*$ to avoid selecting transients.
From left to right we show profiles at $t_* = 9.1 \times 10^3$ yr (a),   $7.49 \times 10^4$ yr (b) and 
$1.036 \times 10^5$ yr (c) after the merger, these being the same snapshots used in Figure 1 and 2 of the main paper. The time
of the first snapshot (a) was indeed chosen near the maximum peak of the $m=2$ mode, which also corresponds to a maximum inflow
rate at scales of $\sim 10-20$ pc.  At time (b) the Jeans collapse has already started at parsec scales, as shown
by the very large inflow rate, which is indeed the highest measured throughout the simulation. At larger ($\sim 10$ pc) 
scales the inflow has instead decreased because the dominant global $m=2$ mode has also weakened considerably
(bottom panel). At time (c)
the $m=2$ mode has faded considerably (higher order modes are now shown but they are even weaker at this time),
the inner region is stable while outside the central parsec a net outflow is seen,
which is responsible for the spreading of the disk highlighted in Figure 2 of the main paper.}

\end{figure}

\paragraph{Star formation and feedback}

The conversion of gas into stars is not included in the refined calculation with particle splitting presented in this Letter,
although its (important) radiative feedback effect on gas thermodynamics is included with the choice of the EOS (see previous subsection).
Indeed the refined calculation is carried out for less than $1\%$ of the starburst duration of $10^8$ yr
indicated by the companion merger simulation carried out without splitting$^{8,9}$, suggesting that the conversion
of gas into stars is not important on the timescales of interest for this work.
Cautionary remarks regarding this point are however necessary. Had we included
star formation directly in the refined simulations we would have probably found local gas consumption timescales shorter
than the global starburst timescale since with splitting much higher densities are resolved and the star formation rate
depends on the local gas density. 
As for the first issue, we can obtain a rough
estimate of how short the star formation timescale can be in the following way. To begin with, in the
nuclear disk most of the gas is at densities above $100$ atoms/cm$^3$. At these densities molecular
hydrogen formation is efficient $^{37}$.
Let us then make the extreme assumption that all the gas in the disk is molecular and
readily available for star formation. Then, let us simply assume that
molecular gas will be turned into stars on the local orbital timescale. Star formation in molecular clouds
is rather inefficient, and typically $\le 30\%$ of the dense, molecular gas is converted into stars, possibly
because internal turbulence in the clouds prevents them from collapsing altogether  $^{35}$. Therefore let us
write the star formation rate in the nuclear disk as a whole as $dM_*/dt = 0.3 \times M_{gas}/T_{orb}$,
where $T_{orb} = 2.5 \times 10^5$ years, the orbital time at the disk half mass radius, $25$ pc, and
$M_{gas} = 3 \times 10^9 M_{\odot}$ is the mass of the enclosed mass of the disk at such distance.
The resulting upper limit on the 
star formation rate is $3600 M_{\odot}$/yr , about 80 times higher than that estimated
in the low-res simulations. Nonetheless, even with such high star formation rate only $10\%$
of the gas within 25-30 pc (this is also the region where the first inflow occurs - see next section
and Figure 1) would be converted into stars during the time required to form the central supermassive cloud ($\sim 10^5$ years).
On the other end, this high star formation rate justifies the use of the EOS valid for a starburst regime 
even on the short timescales probed by our calculations.
These simple estimates suggest that our results should be qualitatively valid in general.

Nonetheless, the arguments just outlined are based on
simple scaling laws and, in particular, neglect the multi-phase nature of the ISM in the nuclear disks, that is important
for star formation and is not accounted for in our EOS model. Furthermore, they are based on using average
properties in the disk. In principle at radii of order a parsec or less higher densities and shorter dynamical
times would yield higher star formation rates, potentially lowering the gas mass available to the
central  cloud.
How much of the cloud mass would be lost to stars will also depend on the balance of heating and cooling {\it within
the cloud}, which is beyond the reach of our resolution. However, as explained in the letter, the high temperature
of the cloud, $T > 10^7$ K, and the fact that further collapse should occur nearly isothermally as $\gamma$ 
approaches unity at even higher densities, thus preserving such high temperatures, argue against the 
importance of star formation within the cloud. Nonetheless, the thermodynamics in the interior of the cloud in
the later stages of the collapse is not necessarily captured by
our EOS model (see $^{38}$ on the importance of the form of the EOS for fragmentation of gas clouds). 
For example, if a quasi-star forms at scales well below our resolution as a result of the collapse, only
a small fraction of its mass would collapse instantaneously into a black hole, while the rest will be
accreted at super-Eddington rates from the gaseous envelope of the quasi-star on timescales of $10^5$
yr$^{39}$
In this case there will be time for the rapidly growing seed black hole to affect the surrounding 
gas via radiative feedback, which would translate in a modification of the gas EOS and in a reduced
efficiency in the growth of the massive seed.
This is why we decided to be conservative
and assumed that only $0.1\%$ of the central gas cloud would end up into a black hole. 

Along with star formation, another astrophysical aspect that we include in a very simplified way is feedback
from star formation. 
Radiative feedback from stars is implicitly included in our choice
of the effective equation of state (see above), but feedback from supernovae
explosions is not taken into account.
However, feedback  from supernovae type II, would contribute to both heating the gas and increasing
its turbulence (the timescale of supernovae type II explosions is sufficiently short to be
relevant here), which should go in the direction of decreasing the star formation rate and therefore
strengthening our previous argument concerning the role of star formation. It would also tend
to stiffen the equation of state even in the central, high density regions, likely bringing our ``mean''
effective $\gamma$ in closer agreement to the values adopted here.
On the other end, a quick calculation suggests that feedback should not have a major impact on the
global energetics of the nuclear disk. In fact, assuming the upper limit on the 
star formation rate of $3600 M_{\odot} / yr$ and 
a Miller-Scalo initial
stellar mass function we obtain that
supernovae should damp $\le 2.5 \times 10^{52}$ erg/yr ($7 \times 10^{48}$ erg per solar mass of stars
formed) into the surrounding gas, corresponding to   $\le 2.5 \times 10^{57}$ erg damped
during the black hole formation timescale, $\sim 10^5$ yr.
This is at most $20\%$ of the the overall internal energy budget of the gas in the nuclear disk
(the sum of turbulent, rotational and thermal energy), hence the effects on thermodynamics
will be  moderate. In addition most of this energy will be deposited outside the inner
few parsecs since there the strong radial inflow will dominate over star
formation because there the inflow timescale is shorter than the orbital timescale on which
star formation proceeds. As a result, the dynamics of the gas inflow, that are crucial in our
formation model, should not be significantly affected.

\paragraph{Feedback from the accreting massive seed black hole}

Finally, once the black hole forms it will start accreting and will lose part of the accretion energy either
radiatively or in part radiatively and in part in form of mechanical energy if a powerful jet is produced as observed
in some of the local and distant AGNs.  This is the so called "AGN feedback", whose effect we have not 
taken
into account in the simple calculation of the timescale for the black hole to grow up to $10^9 M_{\odot}$ . 
Indeed, based on recent calculations$^{40}$, a SMBH starting
with $10^5 M_{\odot}$ should be able to accrete enough mass by $z=6$ even when the radiative feedback from black hole
accretion (AGN feedback) is accounted for.
Note that hydrodynamical effects and momentum deposition due to jets
would deserve a separate discussion but are currently not well understood.

At low redshift two accreting SMBHs should be already present in the galaxies as they merge. Heating of the 
surrounding gas by "AGN feedback" might be strong enough to overcome cooling and produce a stiffer
equation of state in the nuclear gas$^9$. In this case, either a dense nuclear disk will not 
form at all$^9$, and so no multi-scale scale gas 
inflows will occur, or the higher pressure support of the gas within the nuclear disk could prevent the 
secondary inflow and thus the formation of the central supermassive cloud.


\section {Mass transport and stability of the nuclear region}

We have measured the strength of the non-axisymmetric modes in the nuclear disk using a Fourier decomposition in
order to establish a clear correlation between the regions of the disk at which the maximum inflow occurs
in the nuclear disk and the amplitude of the strongest mode. This is shown in the panels of Figure 1.
The strongest mode in the disk is at all times a two-armed spiral, corresponding to $m=2$ in the 
Fourier decomposition, as also apparent in Figure 1 (top panel) of the main paper. Such mode is the
imprint of the collision between the two galaxy cores.
The inflow rate is remarkable, peaking at $> 10^4 M_{\odot}$/yr, which corresponds to radial velocities
of about $100$ km/s. This is sustained for only a few $10^4$ yr, allowing to bring a few $10^8 M_{\odot}$
of gas within the inner 10 pc. Note that the large radial velocities are of order of the turbulent
velocities seen in nuclear disks residing at the center of merger remnants$^{32}$.
In Figure 1 we also show the second radial inflow triggered by the onset of the Jeans collapse of the gas
within the central parsec. The collapse begins at about a parsec scale as soon as the enclosed mass
climbs above the local Jeans mass ($\sim 7 \times 10^7 M_{\odot}$ at $r=1$ pc, note that using the
Bonnor-Ebert mass would yield essentially the same result within a factor $\sim 2$). 
Gas at about a parsec scale rotates at a speed of about $v_{rot} \sim 600$ km/s but it is pressure supported as the temperature raises close
to $10^8$ K at such scales owing to adiabatic compression ($v_s \sim 1000$ km/s $> v_{rot})$. 
The fact that pressure provides the most important
support against gravity justifies our use of the Jeans mass to characterize the phase of collapse,
although a more complete description of the process would involve accounting for the 
effect of rotation and the continued mass flux from the outer region of the disk.

We note that, owing
to the high mass resolution reached after applying particle splitting, the local Jeans length is always
well resolved throughout the disk, in the sense that it is about an order of magnitude larger than either the local SPH
smoothing length or the gravitational softening$^{27}$.
The smoothing length is comparable with the softening at parsec scales, but becomes nearly an order of magnitude
smaller than the latter when approaching a fraction of a parsec, suggesting that, if anything, the collapse may be slowed down
once the cloud contracts to a size of order the softening.
This implies that our conclusion that the inner supermassive
cloud will continue to collapse further should be regarded as conservative; with greater resolution, indeed, not only 
the collapse should continue but should likely be faster.

The right snapshots of Figure 1 show the final expansion of the disk as the spiral arms unwind and transfer 
angular momentum outward of tens of parsecs, generating a net outflow. At this point the inner profile has
reached stability as further collapse is not possible once the resolution limit is reached. Expansion of the
disk as a result of angular momentum transfer driven by spiral modes is a well documented phenomenon in both
gaseous and stellar disks from galactic to planetary scales$^{41,42}$.


In Figure 2 we show the evolution of the Toomre parameter, which strictly measures the {\it local} stability of
a differentially rotating disk to axisymmetric perturbations$^{18}$.
It is important to note that the disk is born out
of equilibrium from the collision of the two cores, rather than becoming unstable starting from an equilibrium
rotational configuration as assumed in the standard perturbative approach.
The disk indeed reaches a near-equilibrium configuration after the phase of intense inflows studied
in this paper. Nevertheless, previous studies on gaseous disks have shown that the Toomre parameter 
provides a good empirical measure of the susceptibility of disks to fragmentation quite irrespective
of how the disk is initially set up, and in this sense also applies well to {\it global} stability to 
generic, non-axisymmetric perturbations$^{43}$.
The Figure shows that the Toomre parameter remains
always in the theoretical regime of stability against fragmentation, although it drops initially to values in the range
$1-1.5$ where strong spiral instabilities are expected and are indeed observed. After the phase of strong non-axisymmetric
instability associated with the inflow and subsequent 
central collapse is terminated (first and second panel) the disk self-regulates to a more stable state, with a minimum $Q \sim 2$.

It is instructive to compare the very high inflow rates measured in the nuclear disk with the expectations
of analytical models that study a self-gravitating, thin isothermal accretion disk in steady state.
Note that such models are based on a local stability analysis, in contrast with the global
character of instability in our nuclear disk.
Under such assumption,
recent works argue that there exists maximum inflow rate in such a disk above which fragmentation,
and thus star formation, will occur$^{44,45}$. In steady state, the maximum inflow rate can be expressed as
$\dot M_{\rm max} = 2 \alpha {{v_s}^3 \over G}$, where $\alpha \sim 0.06$ is the maximum disk viscosity, resulting
from gravitational stresses, $G$ is the gravitational constant,
and $v_s$ is the sound speed. They consider protogalactic disks with temperatures $\sim 4000$ K
($v_s \sim 5$ km/s) for which $\dot M_{\rm max} = 10^{-2} M_{\odot}/yr$. In our disks the thermal sound speed is much higher,
but because the disks are in a gravoturbulent state, it is more sensible to consider the sum of the thermal sound speed and 
turbulent
velocity dispersion, as we have done for the Toomre Q parameter (see Figure 2).
This amounts to about $600-700$ km/s at scales of $25$ pc
(the scale of the first inflow - see Figure 1), which implies an increase of a factor up to $140^3 = 2.7 \times 10^6$, which
would yield a maximum inflow rate $\sim 10^4 M_{\odot}$/yr, quite in agreement with the results shown in Figure 1 (first panel).
The second inflow at parsec scales is even higher and seems to overshoot the maximum rate possible without fragmentation, but 
since this
is triggered by the Jeans collapse of the central cloud rather than by transport by spiral waves (see Figure 1) the argument
based on the maximum viscous stress associated with gravitational instability does not apply anymore. In addition, the
assumption of steady state, which is not appropriate in general under the highly dynamical conditions of our nuclear disk, 
clearly does not apply once the central cloud begins to collapse. 
As anticipated above, the analytical approach just outlined also stems
from a local stability analysis of self-gravitating accretion disks$^{36}$, while the gravoturbulent
state of the nuclear disk is inherently related to the global character of the non-axisymmetric instability; in 
globally unstable disks "effective" $\alpha$ viscosities even larger than unity can easily arise$^{45}$.
Dropping the constraint $\alpha \sim 0.06$,
as suggested by the nature of global instability, would allow even higher inflow rates without fragmentation. 

In any case, our simple analysis shows how a hot, gravoturbulent disk can easily sustain accretion rates orders of magnitude
higher than those of cold, non-fragmenting protogalactic disks considered in recent literature, hence allowing naturally the
formation of central supermassive objects without drainage of gas by star formation. 
These particular conditions arise naturally 
in the highly perturbed nuclear disk emerging from a galaxy merger, and should indeed be interpreted as
a by product of the starburst itself. Thus, in our model star formation has indeed a positive role
in allowing the direct collapse since it provides enough heating to prevent large-scale fragmentation.

Finally,  typical galaxies at high redshift would
have larger reservoirs of gas to form and feed the black hole relative to the
initial conditions adopted in this Letter.
Their more compact galactic
disks (disk sizes are expected to scale as ${(1 + z)}^{-3/2}$ as their embedding dark 
halos in the CDM cosmogony, see also next section) would have produced more compact and dense gaseous
nuclei after the merger, whose shorter dynamical timescales would drive an
even faster mass transport towards the center via gravitational instabilities.
Hence the results presented here should be regarded as conservative for the purpose
of our scenario.


\begin{figure}
\vskip 11.6cm 
{\includegraphics{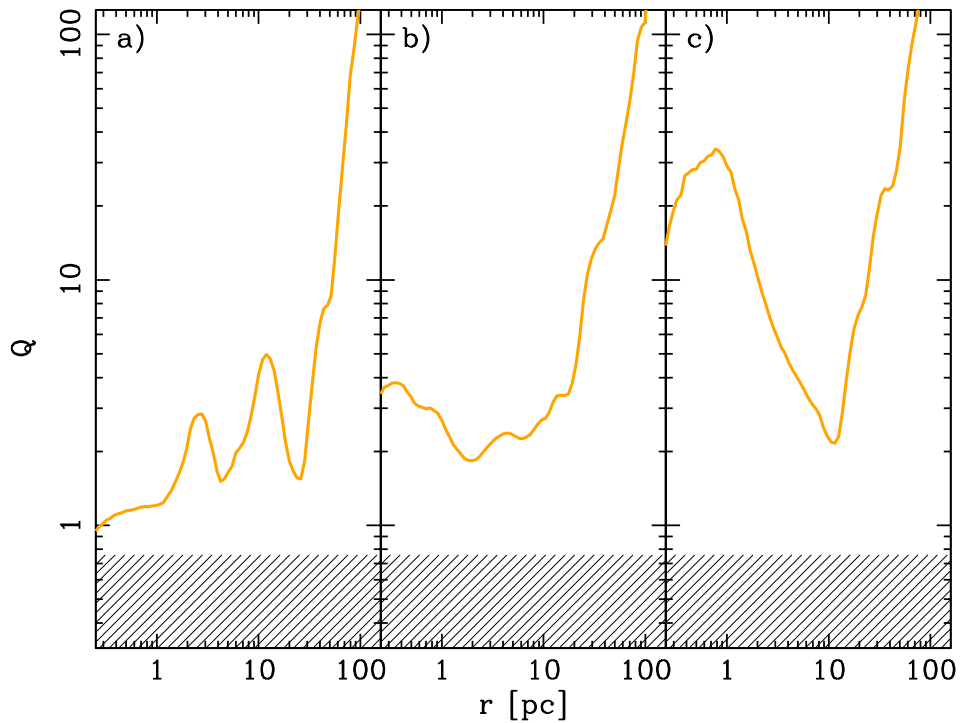}}
\caption[]{\small Azimuthally averaged Toomre Q profile of the disk shown at the same times
as in Figure 1 (time increasing from left to right).
The shaded area corresponds to the instability region marked by $Q \le 0.67$,
this being the stability threshold for finite thickness disks$^{46}$.
The Toomre parameter is calculated as $Q= \kappa v_{s_*}/ \pi G \Sigma$,
where $v_{s_*}$ is the effective sound speed including the contribution of the turbulent (radial)
velocity dispersion, $v_{s_*} = \sqrt {({v_s}^2 + {\sigma_r}^2)}$, where $v_s$ is the thermal sound speed
and $\sigma_r$ the radial velocity dispersion of the gas. Note that $\sigma_r \sim v_s$ outside
the central few parsecs; within the inner few parsecs, especially after the onset of the
Jeans collapse (panels b - c) the thermal pressure is dominant as the gas is strongly
adiabatically compressed (the high central value of $Q$ seen in panel (c) is associated
with the formation of the supermassive hot central cloud).
}
\end{figure}

\vskip 2truecm

\section{Growth of the massive black hole seed}

The time $t(M_{\rm BH})$ required for a black hole of initial mass
$m_{\rm seed}$ to reach a mass $M_{\rm BH}$ assuming Eddington limited accretion$^{47}$ is given by
$t(M_{\rm BH})= \tau \times{f_{\rm edd}}^{-1} \times
({\epsilon_r \over {1 - \epsilon_r}}) \times {\rm ln}(M_{\rm BH}/m_{\rm seed}$), 
where $M_{\rm BH}=10^9 M_{\odot}$, $m_{\rm seed}=10^5 M_{\odot}$, $f_{\rm
  edd}$ expresses at which fraction of the Eddington limit the black
hole is accreting, and $\epsilon_r$ is the radiative efficiency.
In our case $M_{\rm BH}=10^9 M_{\odot}$ and $m_{\rm seed}=2. 6 \times 10^5 M_{\odot}$
(corresponding to $0.1 \%$ of the mass of the collapsing central cloud in our simulation).
Furthermore, we choose standard parameters $f_{\rm edd}=1$, $\epsilon_r=0.1$
and, to compute the characteristic accretion timescale $\tau$,
we adopt a molecular weight per electron for a plasma at zero metallicity with
cosmic abundance of hydrogen ($X=0.75$) and helium ($Y=0.25$), $\mu_e = 1/(1 - Y/2)
= 1.14$, so that $\tau = 0.45 {\mu_e}^{-1}= 0.395$ Gyr. Note that metallicity effects are marginal
in this calculation because they only have a negligible effect on the value of the
molecular weight$^{48}$.
With these choices, we obtain  $t(M_{\rm BH}) = 0.362$ Gyr.
Assuming that the seed black hole can accrete at the Eddington limit
($f_{\rm edd}=1$) is justified by the fact that the hole would accrete the gas
belonging to the nuclear disk, which has very high densities $n_H \sim 10^5-10^8$ atoms/cm$^4$
at scales below 10 pc. Such densities are several orders of magnitude higher
than e.g. those of the gas surrounding the small black hole seeds formed by the
collapse of Pop III stars, which accrete at sub-Eddington rates$^{49}$.
Therefore, not only our model can lead to seeds that are much more massive
relative to those resulting from Pop III stars, but also the following gas
accretion and growth occurs in a much more favourable environment.

Although we have adopted standard values of the parameters to calculate  $t(M_{\rm BH})$ we can
ask how much these can be varied while still being able to reach
a billion solar masses within the first billion years. The timescale,
due to its functional form, is especially sensitive to the radiative 
efficiency. Theoretically,
the actual value of $\epsilon_r$ depends on the spin of the black hole, which
in turn depends on the uncertain mechanism of accretion, such as if the
accretion occurs with or without magnetohydrodynamical (MHD)
effects$^{47}$.
It may be as large as $0.42$ for maximally spinning black holes$^{47}$
accreting in standard thin disks , in which case $t(M_{\rm BH})$ would exceed $1$ Gyr. 
The timescale would instead be $\sim 0.8$ Gyr for $\epsilon_r = 0.2$, as possible in 
MHD disks.
It is important to stress that values of $\epsilon_r$ in the range $0.1-0.2$
should be regarded as more realistic since they are independently inferred 
from the ratio $R$ of the QSO plus AGN luminosity density to the mass density
of SMBHs in nearby galaxies$^{50}$.
In summary,  $t(M_{\rm BH}) < 1 $ Gyr requires $\epsilon_r \le 0.2$ and $f_{\rm edd} \ge 0.7$.
On the other end, if the galaxy merger occurs at $z \sim 8$,
and the black hole has to grow to its final mass by $z \sim 6$, the time available for
accretion is $< 0.5$ Gyr for the standard WMAP5 cosmology, which then favours 
our standard choice of parameters  $f_{\rm edd}=1$ and  $\epsilon_r=0.1$.

As highlighted in the Letter, for the SMBHs to grow as required in the short
time available it is also necessary that the merger timescale is 
significantly shorter than a billion year. Using simple scaling relations for
galaxies and their halos in CDM models$^{51}$ one can show that, at a fixed virial
mass $M_{vir}$, these are both smaller and denser compared to $z=0$. This reflects the fact
that the Universe itself is denser at higher redshift. This reduces significantly
the typical orbital time between pairs of collapsed objects, and thus the merging timescale,
relative to $z=0$. The orbital time
during the merger (assuming for simplicity a circular obit) 
is of order $2 \pi R_{vir}/V_c$, where the halo circular velocity $V_c$ and the virial radius $R_{vir}$
at fixed $M_{vir}$ increase and decrease, respectively, with increasing redshift.
Using the standard scaling relations of Mo, Mao \& White$^{50}$ for such
quantities one can show that $T_{orb} \sim (1 + z)^{-3/2}$. Since at $z=0$ the
merging timescale on a typical cosmological orbit is $\sim 5$ Gyr$^8$, it follows
that at $z=8$ this drops to $\sim 0.2$ Gyr, which is comfortably shorter than a 
billion year.

Finally, in the conventional model in which light seed black holes originate
from the collapse of Pop III stars their growth can be hampered if they are
kicked out after merging with another hole due to the '`gravitational rocket''
effect$^{52}$. This phenomenon can have an important, negative impact on
the growth of  large black holes$^{53}$.
In our direct formation model two obvious scenarios are possible;
either the black hole grows in a galaxy with no pre-existing hole, as we assumed
throughout the paper, or it grows in galaxy where a light ``Pop III seed'' hole
is also present (this would have remained light due to inefficient accretion).
In the first case the gravitational rocket is not important. In the second case, the
lighter black hole could be kicked out of the galaxy, but since it would be orders
of magnitude less massive it will not have an impact on the final mass growth of the
primary black hole formed by direct collapse. In a third, less trivial scenario, 
the galaxy merger remnant 
could merge with a third, similarly massive galaxy in which another black hole has grown
to a similarly large mass, after forming by direct collapse during a previous merger. 
In this case the gravitational rocket may be important since the two holes may have a similar (large) mass when
they merge at the center of the remnant. However, the probability that this may happen before
each black hole has reached a billion solar masses is quite low, since the merger timescale
at $z=6-8$ is not smaller than the characteristic growth timescale of the black hole
calculated with our standard parameters, rather is of the same order (see main paper).
We conclude that,
in general, kicks due to the gravitational rocket should not affect the conclusions
of our work.

\vskip 2truecm

{\bf{References}}

1.  Wadsley, J., Stadel, J., \& Quinn, T., {\it New Astr}., {\bf 9}, 137 (2004)

2. Barnes, J., \& Hut, P., {\it Nature}, {\bf 324}, 446 (1986)


3. Gingold, R.A. \& Monaghan, J.J., {\it Mon. Not. R. Astron. Soc.}, {\bf 181}, 375 (1977)

4. Monaghan, J.J., {\it Annual. Rev. Astron, Astrophys}, {\bf 30}, 543 (1992)

5. Barnes, J., {\it Mon. Not. R. Astron. Soc.},  {\bf 333}, 481 (2002)

6. Springel, V. and Hernquist, L., {\it Mon. Not. R. Astron. Soc.}, {\bf 333}, 649 (2002)

7. Balsara, D.S., {\it J.Comput. Phys.},121, 357 (1995)

8. Kazantzidis, S., Mayer, L., Colpi, M., Madau, P., Debattista, V., Quinn, T., Wadsley, J. \& Moore,  B., {\it Astrophys. J.}, \textbf{623}, L67 (2005)

9.  Mayer, L., Kazantzidis, S., Madau, P., Colpi, M., Quinn, T., \& Wadsley, J., {Rapid Formation of Supermassive Black 
Hole Binaries in Galaxy Mergers with Gas {\it Science}, {\bf 316}, 1874 (2007)

10. Katz, N. {\it Astrophys. J.}, {\textbf 391}, 502 (1992)

11. Governato, F., Mayer, L., Wadsley, J., Gardner, J. P., Willman, B., Hayashi, E., Quinn, T., Stadel, J.\& Lake, G, {\it Astrophys. J.}, {\bf 607}, 688 (2004)

12. Hernquist, L., {\it Astrophys. J. Supp.}, {\bf 86}, 389 (1993)

13. Springel, V. \& White, S.D.M., {\it Mon. Not. R. Astron. Soc.}, {\bf 307}, 162 (1999)

14. Mo, H. J., Mao, S., White, S. D. M., {\it Mon. Not. R. Astron. Soc.}, {\bf 295}, 319 (1998)

15. Navarro, J.~F., Frenk, C.~S. \& White, S.~D.~M., {\it Astrophys. J.}, \textbf{462}, 563 (1996)

16. {Klypin} A.,  {Zhao} H. \&   {Somerville} R.~S., {\it Astrophys. J.}, {\bf 573}, 597 (2002)


17.  Sargent, M.T., The Evolution of the Number Density of Large Disk Galaxies in COSMOS, {\it Astrophys. J. Supp.}, {\bf 162}, 434-455 (2007)

18. Binney, J., \& Tremaine, S., {Galactic Dynamics}, Princeton University Press (1987, 2008)

19. Blumenthal, R.G., Faber, S.M., Flores, R., \& Primack, J.R., {\it Astrophys. J.}, {\textbf 301}, 27 (1986)

20. Law, David R., Steidel, Charles C., Erb, Dawn K., Larkin, James E., Pettini, M., Shapley, A. E., Wright, Shelley A., 2009.
{\it submitted to Astrophys. J.}, 2009arXiv0901.2930L, {2009arXiv0901.2930L}

21. Khochfar, S. \& Burkert, A., {\it Astron. Astrophys.}, \textbf{445}, 403 (2006)

22.  Dotti, M., Colpi, M., Haardt, F., \& Mayer, L., {Supermassive black hole binaries in gaseous and stellar circumnuclear discs: orbital dynamics and gas accretion}, {\it Mon. Not. R. Astron. Soc.},
{\bf 379}, 956-962 (2007)

23. Kaufmann, T., Mayer, L., Wadsley, J., Stadel, J. \& Moore, B., {\it Mon. Not. R. Astron. Soc.},  \textbf{370}, 1612 (2006)

24. Escala, A., Larson, R. B., Coppi, P. S., \& Mardones, D., {\it Astrophys. J.}, \textbf{607}, 765 (2004)


25. Kitsionas, S. \& Whitworth, S., {\it Mon. Not. R. Astron. Soc.}, \textbf{330}, 129 (2002)

26. Steinmetz, M. \& White, S.D.M., {\it Mon. Not. R. Astron. Soc.},  \textbf{288}, 545 (1997)

27. Bate, M. \& Burkert, A., {\it Mon. Not. R. Astron. Soc.}, \textbf{288}, 1060 (1997)

28. Nelson, A.F., {\it Mon. Not. R. Astron. Soc.},  \textbf{373}, 1039 (2006)



29.  Dotti, M., Colpi, M. \& Haardt, F., {\it Mon. Not. R. Astron. Soc.}, \textbf{367}, 103 (2006)


30. Spaans, M. \& Silk, J.,{\it  Astrophys. J}, \textbf{538}, 115 (2000)

31. Klessen, R.S., Spaans, M., Jappsen, A., {\it Mon. Not. R. Astron. Soc.}, \textbf{374}, L29 (2007)

32. Downes, D. \& Solomon, P. M., {\it Astrophys. J.}, \textbf{507}, 615 (1998)

33. Wada, K., {\it Astrophys. J.}, \textbf{559}, L41 (2001)

34. Wada, K. \& Norman, C., {\it Astrophys. J},  \textbf{566}, L21 (2002)

35. Li, Y.,  Mac Low, M.M., \& Klessen, R. S., {Control of Star Formation in Galaxies by Gravitational Instability},
{\it Astrophys. J.}, {\bf 650}, L19-L22 (2005)

36.  Escala, A., {Toward a Comprehensive Fueling-controlled Theory of the Growth of Massive Black 
Holes and Host Spheroids}, {\it Astrophys. J.}, {\bf 671}, 1264 (2007)

37. Schaye, J. {\it Astrophys. J.}, 809, 667 (2004)

38.  Omukai, K., Schneider, R., \& Haiman, Z., {Can Supermassive Black Holes Form in Metal-Enriched High Redshift
Protogalaxies?} {\it Astrophys. J.}, {\bf 686}, 801-814 (2008)

39.  Begelman, M., {Did Supermassive Black Holes Form by Direct Collapse?}, {\it AIPC}, {\bf 990}, 489-493 (2008)

40. Pelupessy, F.I., di Matteo, T., \& Ciardi, B., {How Rapidly Do Supermassive Black Hole ``Seeds'' Grow at Early 
Times?} {\it Astrophys. J.}, {bf 665}, 107 (2007)


41.  Debattista, V. P., Mayer, L., Carollo, C. M., Moore, B., Wadsley, J., Quinn, T., {The Secular Evolution of Disk Structural Parameters}, {\it Astrophys. J.}, {\bf 645}, 209-227 (2006)

42. Mayer, L., Quinn, T., Wadsley, J. \& Stadel, J.,{\it Astrophys. J.}, {\bf 609}, 1045 (2004)

43. Durisen, R. H., Boss, A. P., Mayer, L., Nelson, A. F., Quinn, T., Rice, W. K. M, {Gravitational Instabilities in Gaseous Protoplanetary Disks and Implications for Giant Planet Formation}, Protostars and Planets V, B. Reipurth, D. Jewitt, and K. Keil (eds.), University of Arizona Press, Tucson, 951, 607-622 (2007)

44. Lodato, G, \& Natarayan, P., {Supermassive black hole formation during the assembly of pre-galactic discs},
{\it Mon. Not. R. Astron. Soc.}. {\bf 371}, 1813 (2006)

45. Rice, W. K. M., Lodato, G.\& Armitage, P. J., {Investigating fragmentation conditions in self-gravitating accretion discs}, {\it Mon. Not. R. Astron. Soc.}, {\bf 364},
L56-L60 (2005)

46. Nelson, A.F., Benz, W., Adams, F.C., \& Arnett, D., {Dynamics of Circumstellar Disks}, {\it Astrophys. J.},
{\bf 502}, 342-371 (1998)

47. Shapiro, S.L., {\it Astrophys, J.}, {Spin, Accretion, and the Cosmological Growth of Supermassive Black Holes} 
{\bf 620}, 59-68 (2005)

48. Shapiro, S.L., \& Teukolski, S.A., {Black holes, white dwarfs and neutron stars. The Physics of Compact 
Objects}, Wiley (1985)

49 . Djorgovski, S. G, Volonteri, M., Springel, V., Bromm, V. \& Meylan, G.   {The Origins and the Early Evolution of 
Quasars and Supermassive Black Holes},
Proc. XI Marcel Grossmann Meeting on General Relativity, eds. H. Kleinert, R.T. Jantzen \& R. Ruffini, Singapore: 
World Scientific (Djorgovski, S. G, Volonteri, M., Springel, V., Bromm, V. \& Meylan, G.   {The Origins and the Early Evolution of Quasars and Supermassive Black Holes},
Proc. XI Marcel Grossmann Meeting on General Relativity, eds. H. Kleinert, R.T. Jantzen \& R. Ruffini, Singapore: World Scientific (2008)

50. Soltan, A., {Masses of quasars}, {\it Mon. Not. R. Astron. Soc.}, {\bf 200}, 115-122 (1982)


51. Mo, H, Mao, S., \& White, S.D.M.  {The formation of galactic discs}, {\it Astrophys. J.}, {\bf 295}, 319-336 (1998)

52. Haiman, Z., {\it Astrophys. J.}, {Constraints from Gravitational Recoil on the Growth of Supermassive Black Holes 
at High Redshift}, {\bf 613}, 36 (2004)

53. Volonteri, M., \& Rees, M.J., {Quasars at z=6: The Survival of the Fittest},
{\it Astrophys. J}, {\bf 650}, 669 (2006)

\end{document}